\documentclass[
reprint,
superscriptaddress,
amsmath,amssymb,
aps,
prl
]{revtex4-2}

\usepackage{graphicx}
\usepackage{dcolumn}
\usepackage{bm}

\usepackage[colorlinks=true, allcolors=blue]{hyperref}

\begin{document}

\preprint{APS/123-QED}


\title{Bifurcated Impact of Neutrino Fast Flavor Conversion on Core-collapse Supernovae Informed by Multi-angle Neutrino Radiation Hydrodynamics}

\author{Ryuichiro Akaho}
\affiliation{Faculty of Science and Engineering, Waseda University, 3-4-1 Okubo, Shinjuku, Tokyo 169-8555, Japan}

\author{Hiroki Nagakura}
\affiliation{Division of Science, National Astronomical Observatory of Japan, 2-21-1 Osawa, Mitaka, Tokyo 181-8588, Japan}

\author{Wakana Iwakami}
\affiliation{Faculty of Science and Engineering, Waseda University, 3-4-1 Okubo, Shinjuku, Tokyo 169-8555, Japan}

\author{Shun Furusawa}
\affiliation{College of Science and Engineering, Kanto Gakuin University, 1-50-1 Mutsuurahigashi, Kanazawa-ku, Yokohama, Kanagawa 236-8501, Japan}

\author{Akira Harada}
\affiliation{Interdisciplinary Theoretical and Mathematical Sciences Program (iTHEMS), RIKEN, Wako, Saitama 351-0198, Japan}
\affiliation{National Institute of Technology, Ibaraki College, Hitachinaka 312-8508, Japan}

\author{Hirotada Okawa}
\affiliation{Faculty of Software and Information Technology, Aomori University, Seishincho, Edogawa, Tokyo 134-0087, Japan}

\author{Hideo Matsufuru}
\affiliation{High Energy Accelerator Research Organization, 1-1 Oho, Tsukuba, Ibaraki 305-0801, Japan}

\author{Kohsuke Sumiyoshi}
\affiliation{National Institute of Technology, Numazu College, Ooka 3600, Numazu, Shizuoka 410-8501, Japan}

\author{Shoichi Yamada}
\affiliation{Faculty of Science and Engineering, Waseda University, 3-4-1 Okubo, Shinjuku, Tokyo 169-8555, Japan}

\date{\today}

\begin{abstract}
In this {\it Letter}, we present a compelling and robust argument for the roles of neutrino fast flavor conversion (FFC) in the explosion mechanism of core-collapse supernova (CCSN), combining the {\it multi-angle} FFC subgrid model rooted in quantum kinetic theory with the multi-dimensional four-species Boltzmann neutrino radiation hydrodynamics.
Employing various progenitor masses and the nuclear equations of states, 
we find that the effect of FFC on CCSN explosion is bifurcated depending on the progenitors.
For the lowest-mass progenitor, FFC facilitates the shock revival and enhances the explosion energy, whereas for higher-mass progenitors its impact is inhibitory.
We identify the mass accretion rate as the key determinant governing this bifurcation. 
When the mass accretion rate is low (high), the contribution of FFC to neutrino heating becomes positive (negative), because the heating efficiency enhancement via FFC-driven spectral hardening of electron-type neutrinos dominates over (is outweighed by) the concurrent reduction in neutrino luminosity.
Our results further highlight the limitations of approximate neutrino transport, and demonstrate that a multi-angle treatment is essential for accurately capturing FFC effects; otherwise, FFCs are missed and even generated spuriously.
\end{abstract}

\maketitle


{\em Introduction.}---
Core-collapse supernovae (CCSNe) are the explosions occurring at the end of massive stars' lives (see \cite{Janka2017hsn..book.1095J,Janka2017hsn..book.1575J,Mezzacappa2020LRCA....6....4M,Burrows2021Natur.589...29B,Mezzacappa2023IAUS..362..215M,Boccioli2024Univ...10..148B,Yamada2024PJAB..100..190Y,Suzuki2024PTEP.2024eB101S,Janka2025ARNPS..75..425J} for recent reviews).
The majority of these explosions are supposed to be driven by neutrino-mediated energy transport. 
The collective neutrino oscillations, especially the fast flavor conversion (FFC), constitute one of the uncertainties in the current CCSN theory
\cite{Sawyer2005PhRvD..72d5003S,Duan2006PhRvD..74j5014D,Duan2006PhRvD..74l3004D,Sawyer2009PhRvD..79j5003S}.
Indeed, the impact of FFCs on CCSN remains poorly understood, largely because the occurrence of FFC is determined by the neutrino angular distributions in momentum space, requiring a multi-angle treatment in neutrino transport.

While the angular dependence of the neutrino momentum distribution is crucial for FFC, previous attempts to investigate its influence on CCSN dynamics are all based on the truncated moment method, which is de facto standard but inherently incapable of determining the onset of FFC. 
In previous simulations \cite{Ehring2023PhRvD.107j3034E,Ehring2023PhRvL.131f1401E,Mori2025PASJ..tmp...18M}, the authors resorted to manually prescribing FFC locations by hand. Such approaches have yielded contradictory results, ranging from positive to negative effects depending on the parameters they chose. Thus, no definitive insight into the actual role of FFC was provided. 
Alternatively, other authors attempted to incorporate FFC with the truncated moment method by reconstructing angular distributions from the lower-order moments \cite{Wang2025ApJ...986..153W,Wang2025arXiv251120767W}.
However, the reliability of such reconstruction schemes is highly questionable, as demonstrated in the previous studies \cite{Nagakura2021PhRvD.104f3014N,Nagakura2021PhRvD.104h3025N,Johns2021PhRvD.103l3012J,Cornelius2025PhRvD.112f3004C}.

In this {\em Letter}, we present a suite of CCSN simulations with the multi-angle neutrino transport to address this issue unequivocally. We deploy our Boltzmann radiation hydrodynamics code that implements a FFC subgrid model, the first ever attempt of that kind.
The occurrence of FFC is judged directly from the angular distributions obtained in the simulations, and the resultant neutrino flavor states are determined through quantum-kinetically motivated prescriptions \cite{Nagakura2019ApJ...886..139N,Zaizen2023PhRvD.107j3022Z}, which are implemented through the Bhatnagar-Gross-Krook (BGK) relaxation scheme \cite{Bhatnagar1954PhRv...94..511B,Nagakura2024PhRvD.109h3013N}.
This extended framework of neutrino transport has been thoroughly validated in our previous work \cite{Akaho2025PhRvD.112d3015A}.
Our simulations encompass both successful and failed explosions, spanning a range of progenitor models and three different nuclear equations of state (EOSs).
We demonstrate that FFC has a bifurcated impact on the CCSN explosion and elucidate the underlying physical mechanisms responsible for this bifurcation.

{\em Numerical Setup.}---
We utilize our general relativistic Boltzmann radiation hydrodynamics code \cite{Nagakura2014ApJS..214...16N,Nagakura2017ApJS..229...42N,Nagakura2019ApJ...878..160N,Akaho2021ApJ...909..210A,Akaho2023ApJ...944...60A}, which solves the Boltzmann neutrino transport and hydrodynamics equations simultaneously.
We employ polar coordinates both for the configuration and momentum spaces, and assume 2D axisymmetry for the former.
A multi-dimensional treatment is essential for studying the impact of FFC on CCSN dynamics. The electron lepton number (ELN) crossings, the seed for FFC, in post-shock regions tend to be suppressed in 1D \cite{Nagakura2021PhRvD.104f3014N}.
We incorporate the standard neutrino matter interaction \cite{Bruenn1985ApJS...58..771B}, augmented with the nucleon-nucleon Bremsstrahlung and the neutrino-electron capture on light and heavy nuclei \citep{Langanke2000NuPhA.673..481L,Langanke2003PhRvL..90x1102L,Juodagalvis2010NuPhA.848..454J}, consistent with the composition of our multi-nuclei EOS.
Further details of the neutrino-matter interactions are discussed in \cite{Nagakura2019ApJS..240...38N}.
For the FFC models, we solve for 4-species; electron-type neutrino ($\nu_e$), electron-type anti-neutrino ($\bar\nu_e$), $\mu$-, $\tau$-type neutrino ($\nu_x$) and $\mu$-, $\tau$-type anti-neutrino ($\bar\nu_x$).
For the no-oscillation models, we only solve for 3-species, and do not discriminate $\nu_x$ and $\bar\nu_x$. 

We employ the progenitor models from \cite{Sukhbold2016ApJ...821...38S} with zero-age main sequence masses of $9$, $12$, $16$, $20M_\odot$. 
For the $9M_\odot$ model, simulations with three EOSs are performed; the Furusawa-Togashi EOS based on the variational method (VM EOS) \cite{Furusawa2017JPhG...44i4001F}, the Dirac-Br\"uckner-Hartree-Fock approach \cite{Furusawa2020PTEP.2020a3D05F} (DBHF EOS), and the chiral effective field theory ($\chi$EFT EOS) (Furusawa et al. in preparation). 
For the rest of the progenitors, simulations are all performed with the VM EOS.
Radial grid has $384$ mesh points, which covers the range $r\in[0:2500]\,\mathrm{km}$ (for $9M_\odot$) and $r\in[0:5000]\,\mathrm{km}$ (for the rest of the progenitors). The zenith angle has 128 mesh points covering $\theta\in[0:\pi]$.
Momentum space is gridded as; the zenith angle grid has $10$ mesh points for $\theta_\nu\in[0:\pi]$, the azimuth angle grid has $6$ mesh points for $\phi_\nu\in[0:2\pi]$ and the energy grid has 20 mesh points for $\epsilon\in\left[0:300\right]\,\mathrm{MeV}$. 
For the metric ansatz, we impose the radial gauge polar slicing condition and assume the spherically symmetric metric with the form
$g_{\mu\nu} = \mathrm{diag}\left[-e^{2\Phi(t,r)},\left(1-{2m(t,r)}/{r}\right)^{-1},r^2,r^2\sin^2\theta\right]$, where the functions $\Phi$ and $m$ are calculated from the angle-averaged matter distribution as in \cite{Akaho2025PhRvD.112d3015A}.

The effect of FFC is implemented to the Boltzmann equation with the BGK relaxation scheme \cite{Nagakura2024PhRvD.109h3013N}.
The asymptotic distributions of FFC ($f^\mathrm{as}$) are evaluated from the current distribution $f$ as
\begin{equation}
f_e^\mathrm{as} = \eta f_e+(1-\eta)f_x, \quad
f_x^\mathrm{as} = \frac{1-\eta}{2} f_e+\frac{1+\eta}{2}f_x, \\
\end{equation}
where the subscripts $e$, $x$ denote electron and heavy lepton types, respectively.
The survival probability $\eta$ is calculated as follows.
First, positive and negative part of the electron-heavy-lepton type number (ELN-XLN) are calculated as
\begin{eqnarray}
A&\equiv&\left|\frac{1}{8\pi^3}\int_{\Delta G<0} d(\cos\theta_\nu)d\phi_\nu \Delta G\right|, \\
B&\equiv&\frac{1}{8\pi^3}\int_{\Delta G>0} d(\cos\theta_\nu)d\phi_\nu \Delta G, 
\end{eqnarray}
where $\Delta G$ is the ELN-XLN defined as
\begin{equation}
\Delta G\equiv \int(f_e-\bar f_e - f_x + \bar f_x)\epsilon^2d\epsilon.
\end{equation}
The distribution for antineutrinos is denoted by barred $\bar f$.
The survival probability is chosen to impose flavor equipartition for the minor part of $A$ or $B$, which is motivated by the microscopic quantum kinetic simulation \cite{Zaizen2023PhRvD.107j3022Z,Xiong2023PhRvD.108f3003X}
\begin{equation}
\label{eq:BA1}
\eta = \left\{
\begin{array}{cc}
1-2A/3B & (B\ge A \,\mathrm{and}\, \Delta G\ge0), \\
1-2B/3A & (B<A \,\mathrm{and}\, \Delta G<0), \\
1/3 & (\mathrm{other\,case}). \\
\end{array}
\right.
\end{equation}
The relaxation time is estimated as $\tau_\mathrm{as} \equiv {2\pi}/{\sqrt{AB}}$.

We first perform no-oscillation simulations, and FFC subgrid simulations are started from the time snapshot where the flavor instability (equivalent to ELN crossing) appears behind the post-shock region ($\sim190-220\,\mathrm{ms}$ after bounce for the range of models performed in this study).
Note that the preshock FFC tend to appear earlier than the postshock one, as already shown in the previous studies \cite{Nagakura2021PhRvD.104h3025N,Akaho2024PhRvD.109b3012A}.
However, preshock FFC tends to have small growth rates and very shallow crossings, and is expected to have little effects \cite{Abbar2022JCAP...03..051A} (but also see \cite{Zaizen2021PhRvD.104h3035Z}).
Actually, we also performed FFC simulations started before the appearance of preshock FFC, and confirmed that it has no secular effects, only affecting the shock stochasticity.

{\em Results.}---
\begin{figure}
    \centering
    \includegraphics[width=\linewidth]{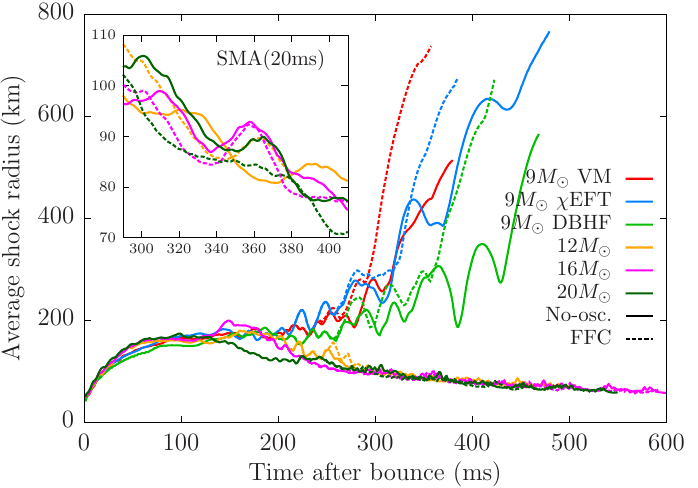}
    \caption{Time evolution of the averaged shock radii. The solid and dashed lines represent no-oscillation and FFC models, respectively. The inset shows the part of the shock evolution for the failed models (simple moving average (SMA) with time window $20\,\mathrm{ms}$ is taken for visibility).}
    \label{fig:shockevo}
\end{figure}
Fig. \ref{fig:shockevo} shows the time evolution of angular-averaged shock radii for all models performed in this study.
First, we briefly review the dynamics of the no-oscillation model.
All $9M_\odot$ progenitor models show shock radii increasing with time, indicating the successful shock revival.
Other progenitor models, $12$, $16$, $20M_\odot$ show maximum shock radii at $\sim100-200\,\mathrm{ms}$ and subsequently contract gradually, indicating the failed explosion.

With FFC, all $9M_\odot$ models show earlier shock expansion compared to the no-oscillation model, and the runaway speed is also larger.
On the other hand, the failed explosion models do not show signs of facilitated shock expansion with the presence of FFC.
Especially, highest ZAMS mass model ($20M_\odot$), which also has the highest accretion rates, shows relatively smaller shock radii with FFC (see the inset in Fig. \ref{fig:shockevo}).

It should be noted that similar bifurcated outcomes have been reported in the previous study \cite{Ehring2023PhRvL.131f1401E}. In that work, however, FFC was implemented through a parametric prescription, and the authors demonstrated that the outcome strongly depends on the manually prescribed parameters.
In contrast, by employing a multi-angle treatment that self-consistently determines the FFC locations and asymptotic states without introducing such free parameters, we demonstrate for the first time that the bifurcation emerges naturally from the progenitor structure (see below for further details).

\begin{figure}
    \centering
    \includegraphics[width=\linewidth]{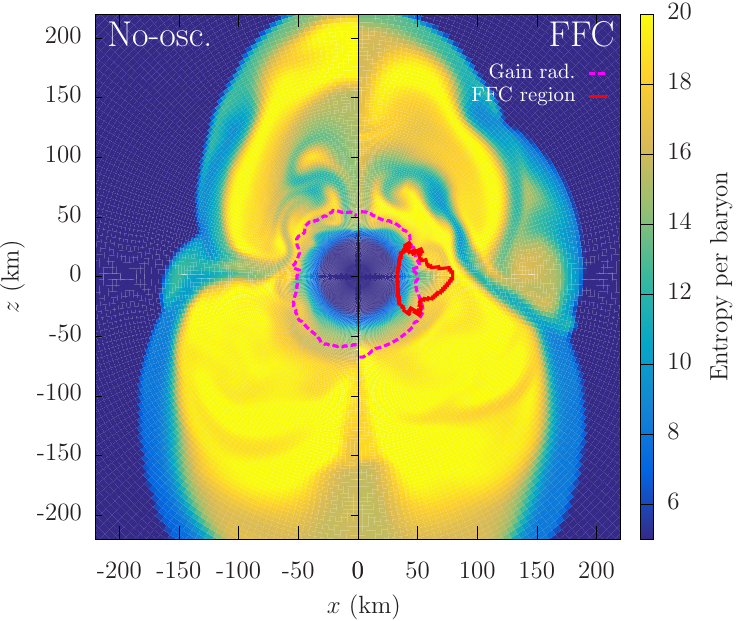}
    \caption{Meridian map of the entropy per baryon for the $9M_\odot$ VM\,EOS model at the $260\,\mathrm{ms}$ after bounce. Left and half panels correspond to no-oscillation and FFC models, respectively. The magenta line represent the gain radius. For the FFC model, outer contour of the FFC region is shown with the red line.}
    \label{fig:FFIregion}
\end{figure}
We have a deeper look into the moment when FFC has initial effect on CCSN dynamics.
Fig. \ref{fig:FFIregion} shows the entropy distribution of $9M_\odot$ VM EOS model at $260\,\mathrm{ms}$ after bounce. This corresponds to $52\,\mathrm{ms}$ after the start of FFC calculation, and the shock begins to show notable deviation from the no-oscillation model.
The FFC region spans across the cooling and gain regions. 
This FFC is typically seen between the neutrino spheres of $\nu_e$ and $\bar\nu_e$, where the latter becomes more forward-peaked than $\nu_e$ \cite{Nagakura2021PhRvD.104h3025N,Harada2022ApJ...924..109H,Akaho2024PhRvD.109b3012A}.
In this model, large-scale convective motion carries low-$Y_e$ proto-neutron star surface matter to the lower latitude, which generates a preferable condition for FFC on the equatorial plane \cite{Nagakura2019ApJ...886..139N,Harada2022ApJ...924..109H}.
The notable feature in Fig. \ref{fig:FFIregion} is that the shock wave of the FFC model clearly shows the expansion in the direction exterior to the FFC region.
This behavior originates from flavor mixing between $\nu_e$ and $\nu_x$, which increases the mean energy of $\nu_e$ and subsequently enhances neutrino heating.
Fig. \ref{fig:FFIregion} is a clear demonstration that FFC in the decoupling region, which typically occurs anisotropically \cite{Harada2022ApJ...924..109H,Akaho2023ApJ...944...60A,Akaho2024PhRvD.109b3012A}, causes anisotropic deformation of the shock wave. Note that the positive effect by FFC spreads laterally at later phases due to the sloshing standing accretion shock instability (SASI), and the angular expansion of the FFC regions. As a result, the spatial correlation between FFC region and shock morphology becomes less distinct at late times.

\begin{figure}
\centering
    \begin{minipage}{\columnwidth}
        \includegraphics[width=\linewidth]{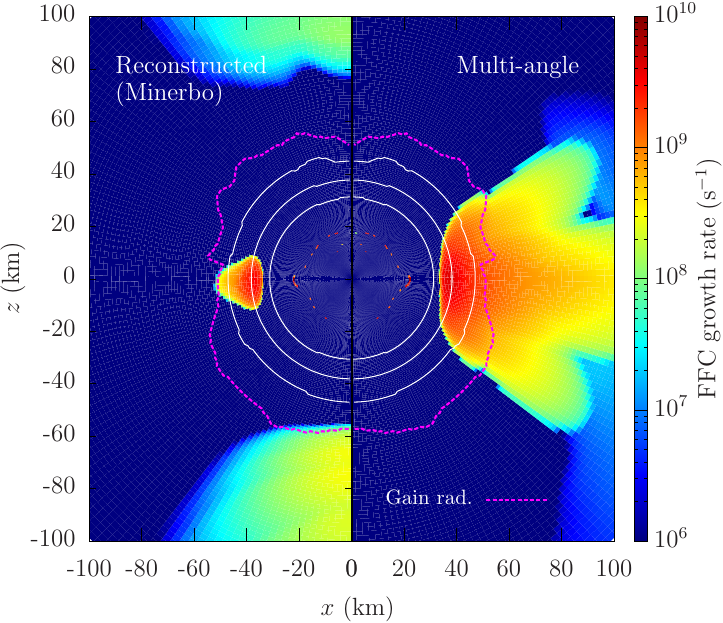}
    \end{minipage} 

    \vspace{0.1cm}

    \begin{minipage}{\columnwidth}
        \includegraphics[width=\linewidth]{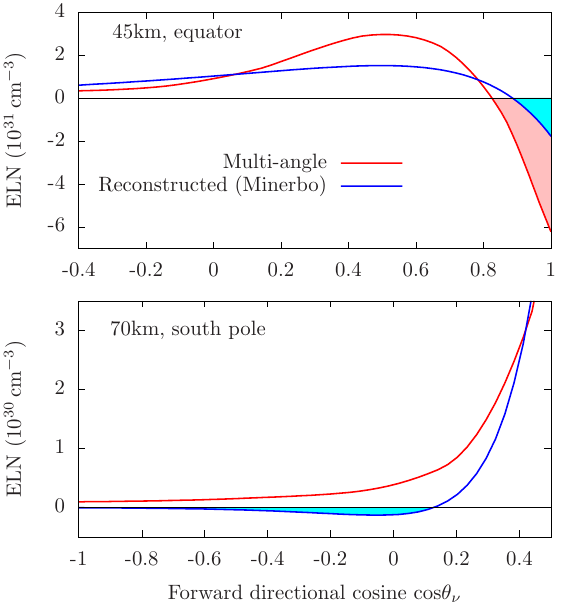}
    \end{minipage}    

    \caption{{\em Top}: Meridian map of the FFC growth rate for the no-oscillation model at $260\,\mathrm{ms}$ for the multi-angle (right) and the one reconstructed with the Minerbo closure (left). The gain radius is shown with the magenta dashed line. The white lines represent the radii which correspond to the density of $10^{10}$, $10^{11}$, $10^{12}\,\mathrm{g}\,\mathrm{cm}^{-3}$. {\em Middle}: Electron-lepton number (ELN) at the $r=45\,\mathrm{km}$ on the equator. Multi-angle and Minerbo cases are compared. The negative ELN region, which is the main driver of FFC, are highlighted with red (multi-angle) and cyan (Minerbo), respectively. {\em Bottom}: same as the middle panel, but for the $r=70\,\mathrm{km}$ on the south pole ($\theta=\pi$).}
    \label{fig:Minerbo}
\end{figure}
A natural question arises as to whether a full multi-angle treatment is truly indispensable, or if moment-based approaches could suffice. To address this, we assess the validity of the FFC detection method used in \cite{Wang2025ApJ...986..153W}, where the maximum entropy closure by Minerbo \cite{Minerbo1978JQSRT..20..541M} was imposed.
In this approach, the distribution is assumed to have the form $f=e^{a\cos\theta_\nu-b}$ and the two parameters $a$ and $b$ are determined from the 0th and 1st moments.

Top panel of Fig. \ref{fig:Minerbo} compares the FFC growth rates at $260\,\mathrm{ms}$ for the no-oscillation $9M_\odot$ VM EOS model, between Minerbo reconstruction and the multi-angle cases. 
It is clear that the decoupling region FFC is mostly overlooked in the reconstructed case, with detections occurring only in a small portion of the cooling region.
As shown in the middle panel of Fig. \ref{fig:Minerbo}, the reconstructed distribution exhibits shallower crossing than the multi-angle case.
This shows that the reconstruction from the moments can underestimate the depth of crossings.
Note that the asymptotic state of FFC is determined by its depth. Therefore, the reconstruction method is likely to underestimate the impact of FFC even in the region where the crossing is detected.
While \cite{Wang2025ApJ...986..153W} concluded that FFC plays only a minor role in CCSN dynamics, it is likely that the presence of FFC and its subsequent effects were artificially suppressed due to the limitations of the reconstruction method.
Furthermore, the growth rate in the reconstructed case shows anomalous FFC region around the poles, which is absent in the multi-angle case. 
The ELN features in this region are depicted in the bottom panel of Fig. \ref{fig:Minerbo}.
Although this region is completely $\nu_e$ dominated over $\bar\nu_e$, the angular reconstruction fails to capture the ingoing direction, thereby artificially inducing a crossing.
Therefore, we conclude that angular reconstruction from lower-order moments can significantly underestimate crossings and also lead to the spurious detection of crossings, which is in line with claims made by other previous studies \cite{Nagakura2021PhRvD.104f3014N,Nagakura2021PhRvD.104h3025N,Johns2021PhRvD.103l3012J,Cornelius2025PhRvD.112f3004C}. We note that, although the contrast between the equatorial and polar directions is artificially enhanced by the axisymmetric assumption, the qualitative trend and our conclusion are expected to hold in 3D. Indeed, similar large-scale anisotropy of FFC region is also observed in 3D simulations (\cite{Nagakura2021PhRvD.104h3025N,Abbar2021PhRvD.103f3033A}).

\begin{figure*}
    \centering
    \includegraphics[width=\linewidth]{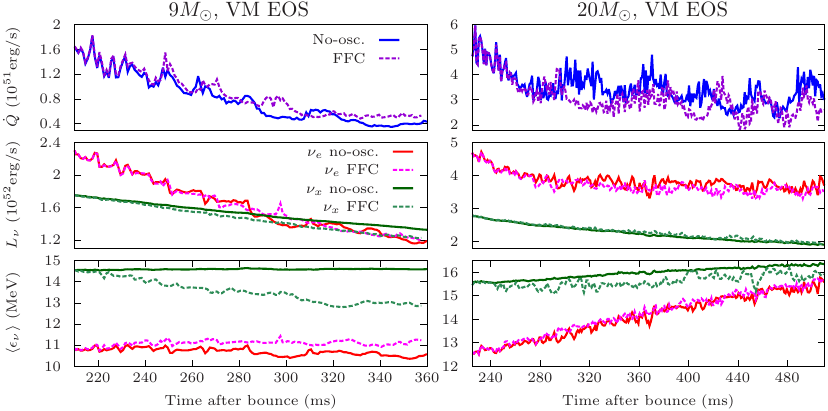}
    \caption{Time evolution of the neutrino heating rates $\dot Q$ (top), neutrino energy luminosities $L_\nu$ (middle), mean neutrino energies $\langle\epsilon_\nu\rangle$ (bottom) for the $9M_\odot$ VM EOS model (left column) and the $20M_\odot$ VM EOS model (right column). The solid and dashed lines denote the no-oscillation and FFC models, respectively.}
    \label{fig:lumi}
\end{figure*}
Finally, we discuss the physical mechanism of bifurcated roles of FFC on shock revival.
Fig. \ref{fig:lumi} compares the heating rates and the properties of the emitted neutrinos. Left ($9M_\odot$) and right ($20M_\odot$) panels are shown to represent successful and failed models, respectively.
As shown in the top panels, the neutrino heating rates tend to be higher with FFC for the $9M_\odot$ model and lower for the $20M_\odot$ model.
The lower two panels compare the energy luminosity and mean neutrino energies, respectively.
Let us first focus on the no-oscillation models (solid lines). 
In the $9M_\odot$ model, $\nu_e$ luminosity decreases with time, and it is eventually overtaken by that of $\nu_x$ at $\sim280\,\mathrm{ms}$.
In the $20M_\odot$ model, on the other hand, $\nu_e$ luminosity stays higher by $\sim2\times10^{52}\mathrm{erg}\,\mathrm{s}^{-1}$ over that of $\nu_x$.
As for the mean energies, the $9M_\odot$ model shows clear separation between $\nu_e$ and $\nu_x$ of $\sim4\,\mathrm{MeV}$, where $20M_\odot$ model shows continuously rising $\nu_e$ mean energy.
This feature is the natural consequence of the accretion rate difference; $9M_\odot$ model has weaker $\nu_e$ emission powered by the accretion, and vice versa (for the time evolution of the accretion rates, see \cite{Burrows2020MNRAS.491.2715B} where the same set of progenitors was used). 

This relative relationships of $\nu_e$ and $\nu_x$ is the primary determinant of the bimodal effect of FFC, as explained below.
FFC effects on CCSN dynamics is the competition of two effects; enhancement of the $\nu_e$ mean energy (positive effect) and decrease of the $\nu_e$ number flux (negative effect).
For the $9M_\odot$ model, the former effect is stronger due to the large separation of mean energies, and the latter is not so dramatic (especially after $L_{\nu_e}\lesssim L_{\nu_x}$). 
For the $20M_\odot$ model, on the other hand, the latter effect is more dramatic due to large separation of luminosity between $\nu_e$ and $\nu_x$, and mean energies being close each other.
This result is also in agreement with previous simulations \cite{Nagakura2023PhRvL.130u1401N,Xiong2025PhRvL.134e1003X,Akaho2025PhRvD.112d3015A}, which reported the suppression of neutrino heating due to FFC. In those works, due to the spherical symmetry, the employed CCSN models feature substantially higher mass accretion rates than the $9M_\odot$ model considered in the present multi-dimensional study.
It should also be noted that the large mass accretion of lepton-rich ($Y_e \sim 0.5$) matter enhances the disparity in number flux between $\nu_e$ and $\bar{\nu}_e$, thereby suppressing the formation of ELN angular crossings (and consequently FFC). This also explains the weak influence of FFC observed in the $20M_\odot$ model.

{\em Summary.}---
We unveiled the role of FFC on CCSN dynamics, by coupling multi-angle FFC modeling with the multi-dimensional neutrino radiation hydrodynamics.
Our study establishes a robust relationship between the accretion rate and the impact of FFC; FFC promotes shock expansion in the models with lower accretion rates, whereas it may inhibit dynamics in high-accretion environments.
We also observed that the asymmetric FFC initially causes asymmetric deformation of shock due to asymmetric neutrino heating, which can be only captured by self-consistently treating the FFC regions.

We also assessed the capability of the truncated-moment approach for FFC subgrid modeling. We found that the zeroth and first angular moments cannot provide enough information to capture most of the angular crossings that usually appear in CCSN. This implies that the multi-angle treatment is mandatory for CCSN modeling with taking into account the FFC effects.

Although our present result is robust, the coarse-grained modeling of FFC’s nonlinear evolution may require further sophistication \cite{Fiorillo2024PhRvL.133v1004F,Urquilla2025arXiv251023917U,Liu2025PhRvD.111b3051L}.
In addition, it is of paramount importance to investigate the role of collisional flavor instability \cite{Johns2023PhRvL.130s1001J}, which occurs at smaller radii than FFC \cite{Akaho2024PhRvD.109b3012A} (but see also \cite{Wang2025PhRvD.112f3039W}).
It can lead to more nontrivial asymptotic states such as the flavor swap \cite{Kato2024PhRvD.109j3009K}.
The implementation of such effects to our multi-dimensional simulation will be reported in the future.

\begin{acknowledgments}
This work used high performance computing resources provided by Fugaku supercomputer at RIKEN, the Wisteria provided by JCAHPC through the HPCI System Research Project (Project ID: 230056, 230204, 230270, 240041, 240079, 240219, 240264, 250006, 250166, 250191, 250226, 250326, JPMXP1020200109, JPMXP1020230406), the FX1000 provided by Nagoya University, Cray XC50 and XD2000 at the National Astronomical Observatory of Japan (NAOJ), the Computing Research Center at the High Energy Accelerator Research Organization (KEK), Japan Lattice Data Grid (JLDG) on Science Information Network (SINET) of National Institute of Informatics (NII), Yukawa Institute of Theoretical Physics.
This work is supported by JSPS KAKENHI Grant Numbers 23K03468, JP24K00632, 25H01273, 25K17399, 25K01006.
R. A. is supported by Waseda University Grant for Special Research Projects (Project number: 2026C-742).
W. I. was supported by Grant-in-Aid for Scientific Research
(24K22885), the FY2024 Research Grant (Women Researchers Support
Program) of the Yamada Science Foundation, and Waseda University Grant
for Special Research Projects (Project number: 2024C-477 and
2025C-139) 
\end{acknowledgments}




\bibliography{apssamp}

\end{document}